\documentclass[reprint, superscriptaddress, aip]{revtex4-2}

\usepackage[normalem]{ulem}
\usepackage{graphicx}
\usepackage{color}
\usepackage{verbatim}
\usepackage{subfigure}
\usepackage{amssymb}
\usepackage{amsmath}
\usepackage{enumitem}
\usepackage{comment}
\usepackage{csquotes}
\MakeOuterQuote{"}
\usepackage{graphicx}
\usepackage{xcolor} 
\usepackage{dcolumn}
\usepackage{bm}
\usepackage{natbib}
\usepackage{lipsum}
\usepackage{multirow}
\usepackage{circledsteps}
\usepackage[hidelinks]{hyperref}
\usepackage[capitalize]{cleveref}

\DeclareMathAlphabet\mathbfcal{OMS}{cmsy}{b}{n}

\usepackage[pangram]{blindtext}
\usepackage{subfigure}
\newcommand*\diff{\mathop{}\!\mathrm{d}}
\def\*#1{\mathbf{#1}}

\begin{document}

\title{A Theory of Localized Excitations in Supercooled Liquids}

\author{Muhammad R. Hasyim}
\email{muhammad$_$hasyim@berkeley.edu}
\affiliation{Department of Chemical and Biomolecular Engineering, University of California, Berkeley, CA, USA}

\author{Kranthi K. Mandadapu}
\email{kranthi@berkeley.edu}
\affiliation{Department of Chemical and Biomolecular Engineering, University of California, Berkeley, CA, USA}
\affiliation{Chemical Sciences Division, Lawrence Berkeley National Laboratory, Berkeley, CA, USA}

\begin{abstract}
A new connection between structure and dynamics in glass-forming liquids is presented. We show how the origin of spatially localized excitations, as defined by dynamical facilitation (DF) theory, can be understood from a structure-based framework. This framework is constructed by associating excitation events in DF theory to hopping events between energy minima in the potential energy landscape (PEL). By reducing the PEL to an equal energy well picture and applying a harmonic approximation, we develop a field theory to describe elastic fluctuations about inherent states, which are energy minimizing configurations of the PEL. We model an excitation as a shear transformation zone (STZ) inducing a localized pure shear deformation onto an inherent state. We connect STZs to T1 transition events that break the elastic bonds holding the local structure of an inherent state. A formula for the excitation energy barrier, denoted as $J_\sigma$, is obtained as a function of inherent-state elastic moduli and radial distribution function. The energy barrier from the current theory is compared to one predicted by the DF theory where good agreement is found in various two-dimensional continuous poly-disperse atomistic models of glass formers. These results strengthen the role of structure and elasticity in driving glassy dynamics through the creation and relaxation of localized excitations.
\end{abstract} 

\maketitle

\section{Introduction}
When liquids are cooled below some onset temperature $T_\mathrm{o}$, microscopic motion slows down dramatically, resulting in a super-Arrhenius increase in equilibrium relaxation times $\tau_\mathrm{eq}$ \cite{Angell2000}. 
In this regime, dynamical heterogeneity emerges at the mesoscale, dividing the liquid into localized mobile regions and extended immobile regions \cite{berthier2011dynamical}. 
To understand these phenomena, two perspectives are commonly used. 
In a structure-based perspective, both $\tau_\mathrm{eq}$ and dynamical heterogeneity are understood from the knowledge of liquid structure \cite{widmer2008irreversible,widmer2009localized,lubchenko2015theory,schoenholz2016structural}.
In a dynamics-based perspective, such as one adopted by dynamical facilitation (DF) theory \cite{Chandler2010, Keys2011}, glassy dynamics is driven by spatially localized regions of particle mobility, known as excitations. Assuming that they relax and emerge by the facilitation of nearby excitations in a hierarchical manner, one can account for the super-Arrhenius increase in relaxation times.

Each perspective predicts different forms for the super-Arrhenius trends in $\tau_\mathrm{eq}$ and thus, it remains an ongoing debate whether a structure- or dynamics-based perspective should be used. In the dynamics-based perspective, however, two open fundamental questions remain: (1) what is the origin of localized excitations? and (2) why should excitations facilitate the relaxation and creation of nearby excitations? In this paper, we answer the first question within a structure-based framework. In particular, we show that key properties of an excitation can be computed from the knowledge of the local structure and elastic properties of inherent states, i.e., energy-minimizing configurations of the potential energy landscape (PEL). 
The elastic signatures and corresponding properties have been invoked in prior studies of supercooled liquids in connection to overall structural relaxation\cite{Dyre2006}, and the characterization of stresses and the displacement fields from inherent states \cite{Lemaitre2014,Chowdhury2016,kapteijns2020nonlinear,rainone2020statistical}. In our work, these elements constitute a central component in understanding the origin of localized excitations in DF theory, as well as the ensuing energy barriers. Before we describe our framework, we review DF theory in the next section and show how it is used to predict the super-Arrhenius trend in equilibrium relaxation times.  

\section{Dynamical Facilitation Theory} \label{sec:dftheory}
In DF theory \cite{Chandler2010,Keys2011}, localized excitations drive glassy dynamics below some onset temperature $T_\mathrm{o}$. These excitations are randomly distributed in space at some concentration $c_\mathrm{eq}=e^{-(\beta-\beta_\mathrm{o}) J_\sigma}$, where $J_\sigma$ is the energy barrier to create an excitation, $\beta= 1/k_\mathrm{B} T$ is inverse temperature, and $\beta_\mathrm{o}=1/k_\mathrm{B} T_\mathrm{o}$. 
Although excitations cannot be probed directly, the theory outlines a procedure to compute $J_\sigma$ from particle trajectories \cite{Keys2011}. The procedure relies on an observable $C_a(t)$, which counts the number of particles that have moved by some magnitude $a$ in some time $t$ given by 
\begin{equation}
C_a(t) = \left\langle \frac{1}{N} \sum_{\alpha=0}^N \Theta\left(|\bar{\* r}_\alpha(t)-\bar{\* r}_\alpha(0)|-a\right) \right\rangle  \label{eq:probe_exc} 
\end{equation}
where $N$ is the number of particles, $\langle \ldots \rangle$ is equilibrium ensemble average, $\bar{\* r}_\alpha(t)$ is the position of the $\alpha$-th particle coarse-grained over a small time window $\delta t$, and $\Theta(x)=1$ if $x > 0$ and zero otherwise. 

At short intermediate timescales, a linear regime exists such that $C_a(t) \sim c_a t$, indicating hopping events being produced at some rate $c_a$. If hopping events are indicators for excitations, then the rate $c_a$ must be Arrhenius
\begin{equation}
c_a(T)  = c_a(T_\mathrm{o}) e^{-(\beta-\beta_\mathrm{o}) J_a} \,    \label{eq:excrate}
\end{equation}
where $c_a(T_\mathrm{o})$ is the rate at $T=T_\mathrm{o}$ and $J_a$ is the energy barrier for observing particle displacements of magnitude $a$. The DF theory also sets $J_a$ when $a=\sigma$, where $\sigma$ is the particle diameter, to be the excitation energy barrier $J_\sigma$, since relaxation is measured from particles displacing a magnitude $\sigma$. Once $J_a$ for all displacement magnitudes $a$ are estimated from the slopes of $-\ln c_a(T)$ vs. $1/T$, one can observe that $J_a$ obeys a logarithmic relation
\begin{equation}
J_a - J_{\sigma} = \gamma J_{\sigma} \ln (a/\sigma)  \label{eq:facilitate}  
\end{equation}
where $\gamma$ is a non-universal constant \cite{Keys2011}.  

\begin{figure}
    \centering
    \includegraphics[width=0.95\linewidth]{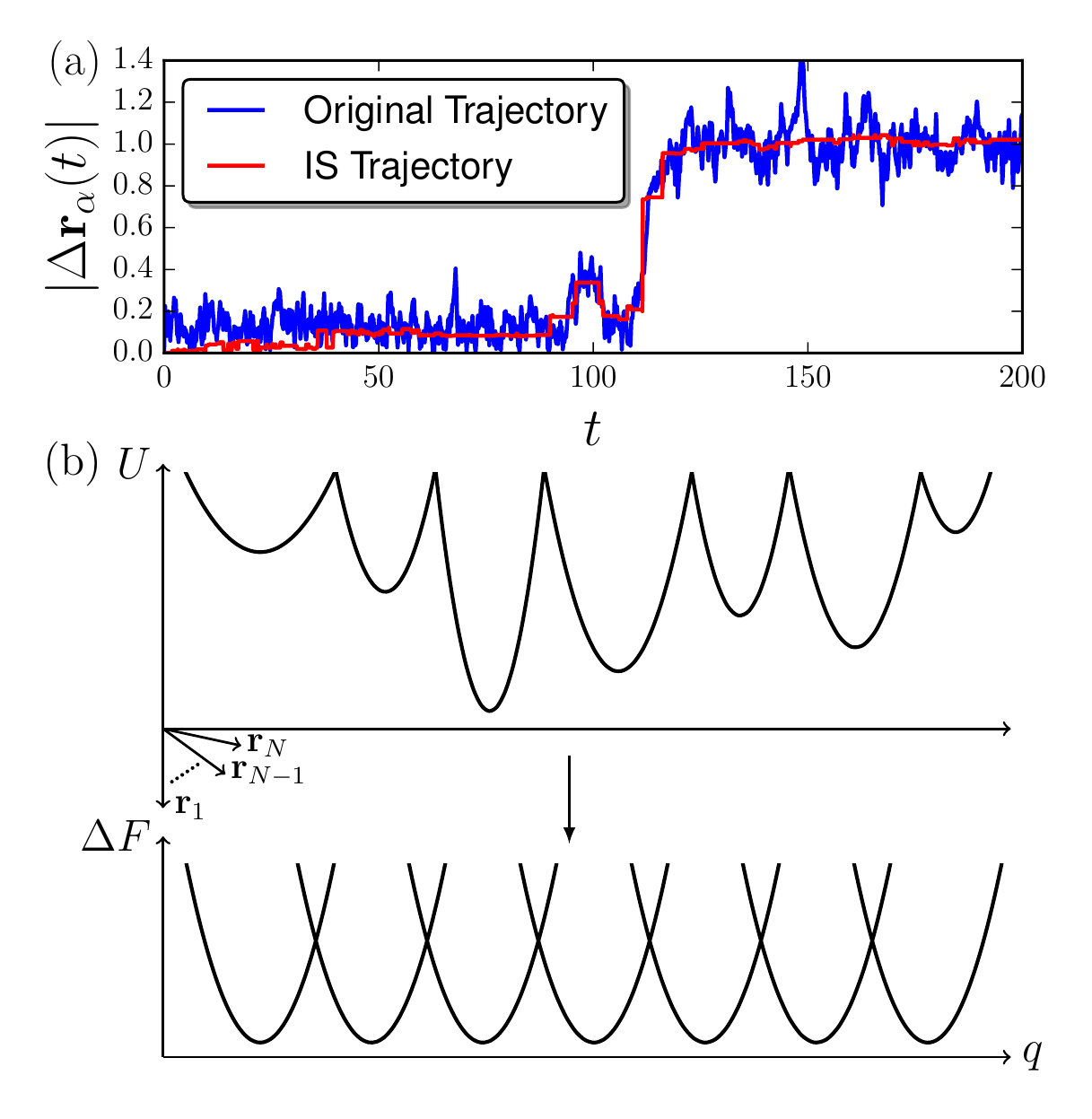}
    \caption{(a) A particle trajectory at temperatures $T < T_\mathrm{o}$ superimposed with its underlying inherent-state (IS) positions showing excitations as hopping events. (b) An illustration of random wells in the PEL being reduced to an equal energy well picture consisting of equivalent neighboring harmonic wells. Here, $q$ denotes the reaction coordinate for the hopping events.}
    \label{fig:general_idea}
\end{figure}

Using Eq.~\eqref{eq:facilitate}, we can turn Eq.~\eqref{eq:excrate} into a power law given by $\frac{c_a}{c_\sigma} \sim \left( \frac{a}{\sigma}\right)^{-\gamma (\beta-\beta_\mathrm{o}) J_\sigma}$. This scaling relation is interpreted as evidence for facilitated dynamics, since the creation of excitations from nearby excitations can translate to motion begetting motion in a self-similar manner \cite{Keys2011}. Guided by kinetically constrained models \cite{Sollich1999,Garrahan2002,Ritort2003}, one can show that facilitated dynamics cascade to create a chain of mobile regions, which terminates at some length $\ell_\sigma=\sigma(1/c_\mathrm{eq})^{1/d_f}$, where $d_f$ is the fractal dimension of heterogeneous dynamics with $d_f \approx 1.8$ and $2.4$ in 2D and 3D respectively \cite{Keys2011}. The energy barrier for equilibrium relaxation $J_{\ell_\sigma}$ is then determined by the same relation as Eq.~\eqref{eq:facilitate} with $\ell_\sigma$ as the new lengthscale, i.e.,
\begin{equation}
J_{\ell_\sigma}-J_\sigma = \gamma J_\sigma \ln (\ell_\sigma/\sigma) \,. \label{eq:facilitate_exc}
\end{equation}

Using Eq.~\eqref{eq:facilitate_exc} along with transition state theory, one can obtain the parabolic law for the equilibrium relaxation time:
\begin{equation}
\ln \left[ \frac{\tau_\mathrm{eq}}{\tau_\mathrm{o}}\right] = \begin{cases}J^2(\beta - \beta_\mathrm{o})^2+(\beta-\beta_\mathrm{o}) E_a & \beta \geq \beta_0 
\\
(\beta-\beta_\mathrm{o}) E_a & \beta < \beta_0 
\end{cases} \label{eq:parablaw}
\end{equation}
where $J=\sqrt{\frac{\gamma}{d_f}}J_\sigma$ is the effective energy scale \cite{Keys2011}. Note that $(\beta-\beta_\mathrm{o}) E_a$ is appended to Eq.~\eqref{eq:parablaw} to accommodate the Arrhenius trend at high temperatures. 

The parabolic form in Eq.~\eqref{eq:parablaw} has been shown to collapse experimental data for relaxation times of a wide variety of single- and multi-component systems \cite{Elmatad2009,Katira2019}. The DF theory has also been used to predict relaxation times of various atomistic systems from molecular simulations \cite{Keys2011, Isobe2016, limmer2013corresponding, takatori2020motility}.
Furthermore, this theory has been used to describe competitions between crystallization and vitrification resulting in the formation of polycrystalline microstructures \cite{hasyim2020theory}. 
However, the two aforementioned fundamental questions regarding the origin of localized excitations and dynamical facilitation still remain to be answered. The answer to the former is provided in this work. In what follows, we describe a general idea towards a quantitative theory of localized excitations, and then proceed to calculate the energy barrier $J_\sigma$ associated with these excitations. 

\section{General Idea} \label{sec:genidea}

To understand the origin of localized excitations, let us first examine how the potential energy landscape (PEL) impacts glassy dynamics \cite{Goldstein1969}. The PEL is a rugged landscape filled with many local minima (Fig.~\ref{fig:general_idea}b), and every energy-minimizing configuration is denoted as an inherent state \cite{stillinger1982hidden}. In the deeply supercooled regime, dynamics proceeds through fluctuations around an inherent state, followed by hopping to the next inherent state \cite{Schroder2000,Keys2011, heuer2008exploring}. As a result,  one may associate every configuration $\{ \* r^\alpha \}$ in the liquid-state trajectory with  a corresponding inherent state $\{ \* R^\alpha \}$ obtained through local energy minimization of $\{ \* r^\alpha \}$ \cite{Bitzek2006} (see Fig.~\ref{fig:general_idea}a). 

Based on these observations, a theory of localized excitations can be constructed by associating the hopping events in the PEL and corresponding transition states to excitations in the DF theory. 
However, the energy barrier $J_\sigma$ in the DF theory obtained from Eq.~\eqref{eq:excrate} corresponds to an average over all individual hopping events. Such an average picture may be obtained by reducing the random energy wells in the PEL to an equal energy well picture in a reaction coordinate space (see Fig.~\ref{fig:general_idea}b). 
With the reduced energy well picture, one can obtain the excitation barrier $J_\sigma$ by studying the barrier-crossing event in this new space. 
The construction of such an equivalent description, leading to an analytical formula for the barrier $J_\sigma$ as a function of key structural properties, consists of the following four steps:

\begin{enumerate}
    \item The first is to develop a field theory for describing fluctuations about inherent states (Sec.~\ref{sec:elasticity} and SM \cite{Note1}, Sec.~1). We show that these fluctuations are governed by an elastic strain energy functional averaged over the inherent states. The corresponding elastic constants can be computed directly from inherent state configurations $\{ \* R^\alpha\}$, thereby connecting the field theory with the particle picture. 
    \item We then model the transition state corresponding to a hopping event as a shear transformation zone (STZ), defined as a pair of force dipoles inducing a localized pure shear (Sec.~\ref{sec:ratetheory} and SM \cite{Note1}, Sec.~2). Using transition state theory (TST) and the elastic field theory, we obtain an  analytical formula for $J_\sigma$ as a function of the elastic constants and the magnitude of the force dipole $f^\ddagger$, which still needs to be determined.
    \item To determine the force magnitude $f^\ddagger$, we model the STZ as a T1 transition event (Sec.~\ref{sec:t1transition} and SM \cite{Note1}, Sec.~3.1), typically studied in the context of cellular re-arrangements\cite{tewari1999statistics,weaire2001physics,cantat2013foams} and also invoked in the studies of glassy dynamics \cite{Eckmann_2008}. The T1 transition state allows us to compute $f^\ddagger$ from the local shear strain $\epsilon_c$ inside the STZ, referred to as the eigenstrain. This strain is a function of the displacement of a particle $u^\ddagger$ participating in the T1 transition event.  
    \item Finally, we use the knowledge of inherent-state local structure to set $u^\ddagger$ as the minimum displacement needed to break an elastic bond involved in the T1 transition event (Sec.~\ref{sec:eigenstrain} and SM \cite{Note1}, Sec.~3.2). Assuming that the bond-breaking event is determined by reorganization of the first solvation shell, we calculate  $u^\ddagger$ from the peaks of the inherent-state radial distribution function (RDF).
\end{enumerate}

These steps constitute a complete construction of a quantitative theory for the barrier, which is then tested on various 2D atomistic continuous poly-disperse models \cite{Ninarello2017} (Sec.~\ref{sec:results}). We focus our attention to 2D systems where we model the bond-breaking events as T1 transitions, and leave the  investigation in 3D for future work. Furthermore, it has been shown that the configurational entropy \cite{ozawa2018configurational} of 2D poly-disperse systems vanishes at zero temperature, allowing us to disregard the possibility of a thermodynamic singularity in relaxation times at finite temperatures \cite{Berthier2019}.

\section{Elastic Strain Energy of Inherent States} \label{sec:elasticity}
We begin by constructing the equal energy well picture by developing a field theory of fluctuations about inherent states.
To study these fluctuations, let us write the canonical partition function $Z$ for $N$-many particles in $d$-dimensions as
\begin{equation}
Z = \frac{1}{\lambda^{Nd} N!} \int \prod_{\alpha=1}^N \diff^d \* r^\alpha \ e^{-\beta U(\{ \* r^\alpha\})} \, \label{eq:partitionfunc}
\end{equation}
where $\lambda$ is the thermal de Broglie wavelength and $U(\{ \* r^\alpha\})$ is the potential energy given by a pair-wise sum of pair potentials $\phi(r^{\alpha \beta})$, i.e., $
U(\{ \* r^\alpha\}) = \sum_{ \alpha, \beta } \phi(r^{\alpha \beta})$ and  $r^{\alpha \beta} = | \* r^\alpha -\* r^\beta|$ is the pair distance between $\alpha$-th and $\beta$-th particle. 

Fluctuations about an inherent state can be introduced into $Z$ by rewriting the phase-space integral into a sum of integrals, each of which is defined over a local region $\mathcal{P}(\{ \* R^\alpha \})$ centered at some inherent state $\{ \* R^\alpha \}$,
\begin{equation}
Z = \frac{1}{\lambda^{Nd} N!} \sum_{\{ \* R^\alpha \} \in \mathcal{C}} \int_{\mathcal{P}(\{ \* R^\alpha \})} \prod_{\alpha=1}^N \diff^d \* r^\alpha \ e^{-\beta U(\{ \* r^\alpha\})} \label{eq:partitionfunc_sw}
\end{equation}
where $\mathcal{C}$ is the collection of all inherent states \cite{stillinger1982hidden}. 

Let us decompose $U(\{ \* r^\alpha\})$ into two parts,
\begin{equation}
U(\{ \* r^\alpha \}; \{ \* R^\alpha \})=U(\{ \* R^\alpha \})+\Delta U(\{ \* r^\alpha \}; \{ \* R^\alpha \}) \label{eq:newpotentialenergy}  
\end{equation}
where $U(\{ \* R^\alpha \})$ is the inherent-state energy and $\Delta U(\{ \* r^\alpha \}; \{ \* R^\alpha \})$ contains both harmonic and anharmonic interactions. Equation~\eqref{eq:newpotentialenergy} allows us to write Eq.~\eqref{eq:partitionfunc_sw} as a product of two partition functions,
\begin{gather}
Z = \frac{1}{\lambda^{Nd} N!} Q_\mathrm{IS} \overline{Q}_\mathrm{fln}\,,  \label{eq:partitionfuncprod}
\\
Q_\mathrm{IS} =\sum_{\{ \* R^\alpha \} \in \mathcal{C}} e^{-\beta U(\{ \* R^\alpha \})}\,,
\\
\overline{Q}_\mathrm{fln} =  \left\langle \int_{\mathcal{P}(\{ \* R^\alpha \})} \prod_{\alpha=1}^N \diff^d \* r^\alpha \ e^{-\beta \Delta U(\{ \* r^\alpha \}; \{ \* R^\alpha \})} \right\rangle_\mathrm{IS}  \label{eq:partfunc_vib}
\end{gather}
where $\langle \ldots \rangle_\mathrm{IS} = \frac{1}{Q_\mathrm{IS}} \sum_{\{ \* R^\alpha \} \in \mathcal{C}} \ldots e^{-\beta U(\{ \* R^\alpha \})}$ is an inherent-state ensemble average; see SM \cite{Note1}, Sec.~1.1 for a detailed derivation of Eqs.~\eqref{eq:partitionfuncprod}-\eqref{eq:partfunc_vib}.
Here, $Q_\mathrm{IS}$ is a partition function for an ensemble of inherent states and $\overline{Q}_\mathrm{fln}$ is a partition function corresponding to fluctuations about those inherent states. 

At lower temperatures, one may invoke the harmonic approximation for $\Delta U(\{ \* r^\alpha \}; \{ \* R^\alpha \})$, and express the energy as a function of particle displacement $\* u^{\alpha} = \* r^\alpha - \* R^\alpha$. This harmonic approximation allows us to rewrite the Boltzmann factor in Eq.~\eqref{eq:partfunc_vib} in terms of a mixture of Gaussian distributions, which further allows us to replace the integration domain $\mathcal{P}(\{ \* R^\alpha \})$ with the full phase space leading to
\begin{align}
\overline{Q}_\mathrm{fln} &\approx \left\langle \int \prod_{\alpha=1}^N \diff^d \* u^\alpha \   e^{-\beta \Delta U(\{ \* u^\alpha \}; \{ \* R^\alpha \})} \right\rangle_\mathrm{IS} \label{eq:partfunc_firstapprox} \\
& \approx  \int \prod_{\alpha=1}^N \diff^d \* u^\alpha \ \left\langle  e^{-\beta \Delta U(\{ \* u^\alpha \}; \{ \* R^\alpha \})} \right\rangle_\mathrm{IS} \,. \label{eq:partfunc_approx}
\end{align}

To arrive at a field theory, the energy $\Delta U(\{ \* u^\alpha \}; \{ \* R^\alpha \})$ can be equivalently represented in terms of the strain tensor field $\epsilon_{ij}= \frac{1}{2} \left[u_{i,j}+u_{j,i}\right]$ where $u_i(\* x)$ is the displacement field. In this representation, the harmonic expansion of $\Delta U$ in index notation can be written as an elastic strain energy functional given by 
\begin{equation}
\Delta U[\epsilon_{ij}; \{ \* R^\alpha \}] \approx \frac{1}{2}\int \diff^d \* x \ \epsilon_{ij} C_{ijkl}(\{ \* R^\alpha \}) \epsilon_{kl} \label{eq:elasticstrain}
\end{equation}
where $C_{ijkl}(\{ \* R^\alpha \})$ is an inherent state elasticity tensor;  see SM \cite{Note1}, Sec.~1.1 for a complete treatment of the expansion. In terms of the field representation, the partition function $\overline{Q}_\mathrm{fln}$ in Eq.~\eqref{eq:partfunc_approx} then becomes  
\begin{equation}
\overline{Q}_\mathrm{fln} \approx \int \mathcal{D} \* u \ \left\langle e^{-\beta \Delta U [\epsilon_{ij}; \{ \* R^\alpha\}]} \right\rangle_\mathrm{IS} \,  \label{eq:partfuncfield-0}
\end{equation}
where $\mathcal{D} \* u$ is the functional measure.

Equation~\eqref{eq:partfuncfield-0} can be equivalently expressed as the following functional integral
\begin{equation}
\overline{Q}_\mathrm{fln} \approx \int \mathcal{D} \* u \ e^{-\beta \Delta F[\* u]}   \label{eq:partfuncfield}
\end{equation}
where $\Delta F[\* u]$ is the effective Hamiltonian given by 
$\Delta F[\epsilon_{ij}] \equiv -k_\mathrm{B} T \ln \langle e^{-\beta \Delta U[\epsilon_{ij}; \{ \* R^\alpha \}]} \rangle_\mathrm{IS}$.
Since $\* u(\* x)$ is integrated irrespective of the choice of $\{ \* R^\alpha \}$, a series expansion for small strains around $\epsilon_{ij} = 0$ (see SM \cite{Note1}, Sec.~1.1) allows us to further approximate $\Delta F[\epsilon_{ij}]$ as
\begin{gather}
\Delta F[\epsilon_{ij}] \approx \frac{1}{2} \int \diff^d \* x \ \epsilon_{ij} C^\mathrm{IS}_{ijkl} \epsilon_{kl}  \label{eq:renormelastic}
\end{gather}
where $C^\mathrm{IS}_{ijkl} = \left\langle C_{ijkl}(\{ \* R^\alpha \}) \right\rangle_\mathrm{IS}$ is the inherent-state ensemble averaged elasticity tensor. Altogether, Eqs.~\eqref{eq:partfuncfield} and \eqref{eq:renormelastic}  form the Gaussian field theory of elastic fluctuations about inherent states, with $\Delta F[\epsilon_{ij}]$ providing the equal energy well picture.

The inherent state elastic constants $C_{ijkl}( \{ \* R^\alpha \})$ can be expressed as a sum of two contributions 
\begin{gather}
C_{ijkl} = C_{ijkl}^\mathrm{B}+C_{ijkl}^\mathrm{NA}\,, \label{eq:elasticitytensor}
\\
C_{ijkl}^\mathrm{B} = \frac{1}{V} \left[\sum_{ \alpha , \beta } \left( \phi_{rr}^{\alpha \beta} R^{\alpha \beta} -\phi_r^{\alpha\beta}\right) \frac{R^{\alpha \beta}_i R^{\alpha \beta}_j  R^{\alpha \beta}_k R^{\alpha \beta}_l}{\left(R^{\alpha \beta}\right)^3} \right]\,, \label{eq:borntensor}
\\
C_{ijkl}^\mathrm{NA} = -\frac{1}{V} \left[ \Xi^\alpha_{ijm} (H_{mn}^{\alpha\beta})^+ \Xi^\beta_{kln} \right]\,, \label{eq:nonaffinetensor}
\\
\Xi^\alpha_{ijm} = \sum_{\alpha \neq \gamma}  \left(\phi_{rr}^{\alpha \gamma}R^{\alpha \gamma}-\phi_r^{\alpha\gamma}\right)  \frac{R_i^{\alpha \gamma}  R_j^{\alpha \gamma} R_k^{\alpha \gamma}}{\left(R^{\alpha \gamma}\right)^3} \label{eq:mismatchforce}
\end{gather}
where $C_{ijkl}^\mathrm{B}$ and $C_{ijkl}^\mathrm{NA}$ are the Born and non-affine contributions to the elasticity tensor, $\phi_{r}^{\alpha \beta}$ and $\phi_{rr}^{\alpha \beta}$ are the first- and second-derivatives of the pair potential at $r=R^{\alpha \beta}$, and $(H_{mn}^{\alpha\beta})^+$ are components of the pseudo-inverse of the Hessian matrix. For a complete derivation of Eqs.~\eqref{eq:elasticitytensor} to \eqref{eq:mismatchforce}, see SM \cite{Note1}, Sec.~1.2.  

The elastic stress tensor $T_{ij}$ for a displacement fluctuation about an inherent state is given by 
$T_{ij} \equiv \frac{\delta \Delta F[\epsilon_{ij}]}{\delta \epsilon_{ij}} = C_{ijkl} ^\mathrm{IS} \epsilon_{kl}$. Noting that $ C_{ijkl}^\mathrm{IS}$ is an averaged property of the inherent state ensemble, we expect it be an isotropic tensor, which for 2D system is given by  $C_{ijkl}^\mathrm{IS} = B^\mathrm{IS} \delta_{ij} \delta_{kl} +G^\mathrm{IS} (\delta_{ik} \delta_{jl}+\delta_{il} \delta_{jk}-\delta_{ij} \delta_{kl})$ where $G^\mathrm{IS}$ and $B^\mathrm{IS}$ are the inherent-state shear and bulk moduli respectively. The 2D effective strain energy functional in Eq.~\eqref{eq:renormelastic} can be recast in terms of the stress tensor as
\begin{equation}
\Delta F =  \int \diff^2 \* x \left( \frac{1}{2B^\mathrm{IS}} (T_{1})^2+\frac{1}{2 G^\mathrm{IS}} \left[ (T_2)^2+(T_3)^2 \right] \right) \label{eq:2delasticenergy}   
\end{equation}
where $T_1 = -\frac{1}{2}\left( T_{xx}+T_{yy}\right)$, $T_2 = \frac{1}{2}\left( T_{yy}-T_{xx}\right)$, and $T_3 = T_{xy}$ (SM \cite{Note1}, Sec.~1.1). Equation~\eqref{eq:2delasticenergy} forms the basis for computing the excitation energy barrier $J_\sigma$ in the next section.

\section{Rate Theory for Elastic Dipoles} \label{sec:ratetheory}

Since we model excitations in DF theory as the barrier-crossing events between inherent states, the rate of such excitations $k_\mathrm{exc}(T)$ can be computed from TST \cite{chandler1978statistical,peters2017reaction}. Let $q(\{ \* r_\alpha \})$ be a reaction coordinate that tracks the progress of a transition pathway connecting one inherent state to the next inherent state. The surface spanned by $q(\{ \* r_\alpha \}) = q^\ddagger$ in phase space delineates the energy basin of one inherent state from the other and thus, defines the transition state. Assuming equilibria between the inherent and transition state, one can compute $k_\mathrm{exc}(T)$ in terms of ensemble-averaged properties as
\begin{equation}
k_\mathrm{exc}(T) = \nu e^{-\beta \Delta F^\ddagger(T)}\,.
\end{equation}
Here, $\nu \equiv  \frac{1}{2}\left\langle \dot{q}(\{ \* r_\alpha \}) \right\rangle_\ddagger$ is the frequency prefactor with $\left\langle \dot{q}(\{ \* r_\alpha \}) \right\rangle_\ddagger$ being the average rate of $q(\{ \* r_\alpha \})$ when the system is at the transition state, and  $\Delta F^\ddagger(T)$ is the transition-state energy barrier, which can be estimated from the elastic strain energy given in Eq.~\eqref{eq:2delasticenergy}.  

If the barrier crossing events correspond to excitations then the rate $k_\mathrm{exc}$ must be proportional to the equilibrium concentration of excitations, i.e., 
\begin{equation}
\frac{k_\mathrm{exc}(T)}{\nu} = e^{-\beta \Delta F^\ddagger(T)} \sim  \frac{c_\sigma(T)}{c_\sigma(T_\mathrm{o})} = e^{-\beta J_\sigma} \,,
\end{equation}
which implies that $\Delta F^\ddagger(T)$ should be at most linear with respect to temperature, i.e., $\Delta F^\ddagger(T)=a+ b T$ where $a$ and $b$ are constants. The validity of such an observation is tested later in the atomistic models.
If this is true, then $J_\sigma$ corresponds to the zero-temperature limit of $\Delta F^\ddagger(T)$:
\begin{equation}
J_\sigma = \lim_{T \to 0} \Delta F^\ddagger(T) \,. \label{eq:jsigmalimit}
\end{equation}
To compute $\Delta F^\ddagger(T)$, one must understand the energetic cost of moving away from any inherent state sampled at thermal equilibrium. Such energetic cost is quantified by the elastic strain energy $\Delta F$ given in Eq.~\eqref{eq:2delasticenergy}. Upon choosing $\Delta F$ as the basis for computing $\Delta F^\ddagger(T)$, we effectively treat the transition state as an elastic mode that brings the system towards the nearest saddle point.

\begin{figure}[t]
    \centering
    \includegraphics[width=\linewidth]{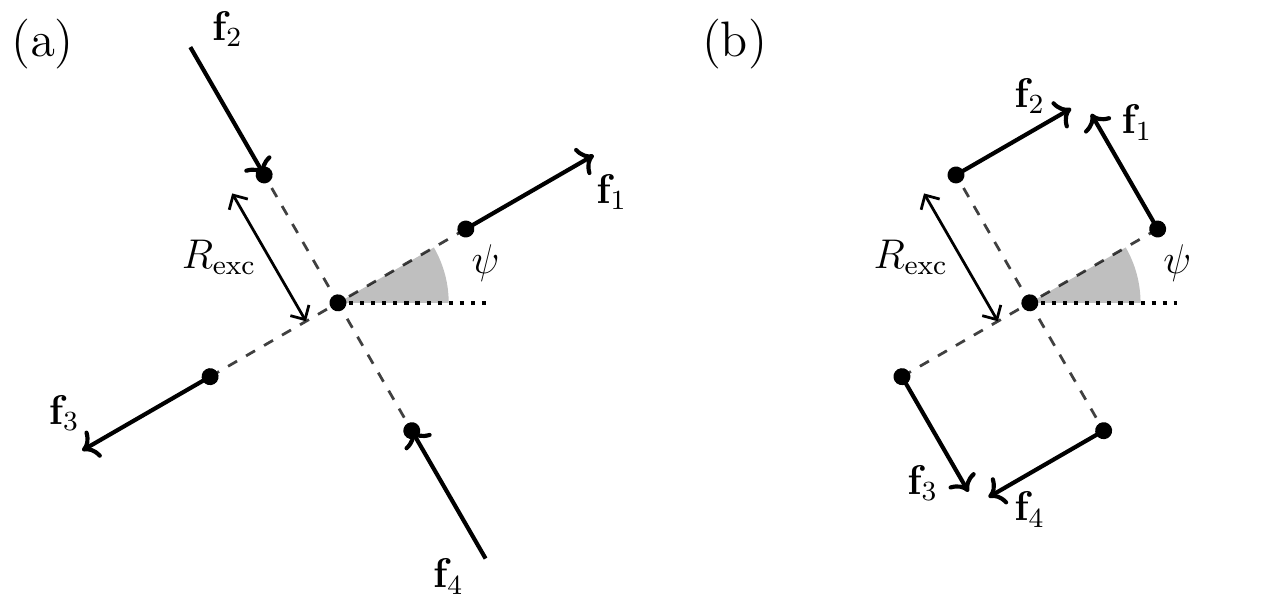}
    \caption{An illustration of two force dipole configurations that produce a state of pure shear, where the point forces act  either parallel (a) or perpendicular (b) to the position vectors with respect to the center of the excitation. The magnitude of all point forces $|\* f_i|=f^\ddagger$ is the same in both configurations making them force and moment free. The remainder of our derivations are based on the configuration in Fig.~\ref{fig:dipoleconfig}a.}
    \label{fig:dipoleconfig}
\end{figure}

Guided by previous studies showcasing elastic signatures of supercooled liquids \cite{Lemaitre2014,Chowdhury2016}, we model the elastic mode as a shear transformation zone (STZ), defined as a localized inelastic pure shear driven by a configuration of force-dipoles. To this end, two simple but equivalent configurations, which are overall force and moment free, are shown in Fig.~\ref{fig:dipoleconfig}. They are comprised of point forces of magnitude $f^\ddagger$ applied to a core region of radius $R_\mathrm{exc}$. 
The difference in these configurations lies only in the orientation of their shear deformations; however, as we will later see in Sec.~\ref{sec:t1transition}, the configuration in Fig.~\ref{fig:dipoleconfig}a is more relevant for a T1 transition event.

\begin{figure*}[t]
    \centering
    \includegraphics[ width=0.825\linewidth]{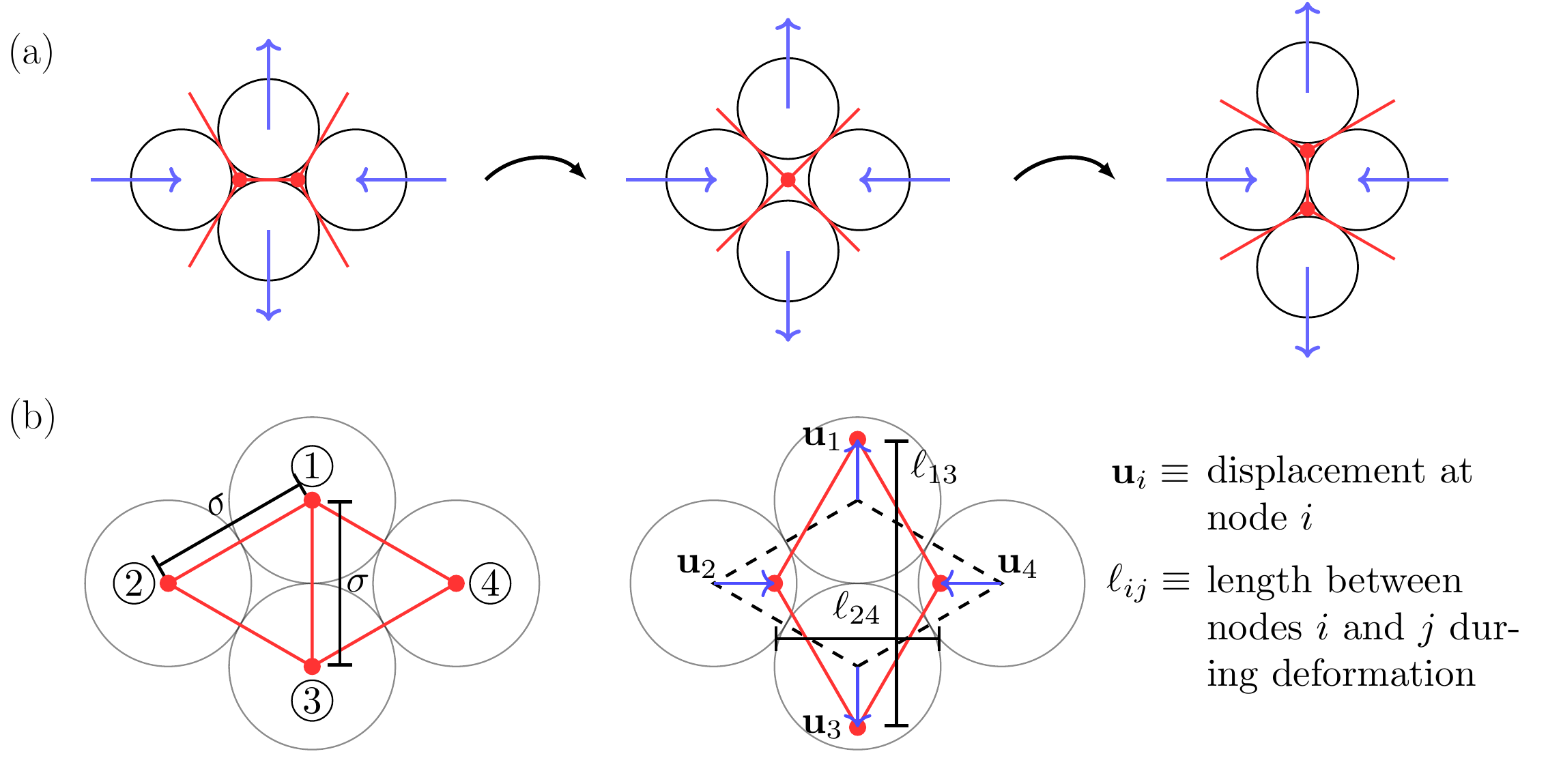}
    \caption{(a) An illustration of the T1 transition represented by a four-particle configuration, the edges of the Voronoi network (red lines), and the forces (blue arrows) involved in the shear transformation. (b) An illustration of the triangulated cell making up the four-particle configuration before (left) and during (right) deformation.}
    \label{fig:t1_transition}
\end{figure*}

The energetic cost to form an STZ is determined by the elastic deformations that the force-dipole configurations impose on the inherent state. Using the method of elastic Green's functions \cite{Balluffi2012}, the stresses corresponding to the configuration in Fig.~\ref{fig:dipoleconfig}a  in polar coordinates $(r,\theta)$ for an orientation angle of $\psi=0$ are given by 
\begin{align}
T_1^\ddagger(r,\theta) &= f^\ddagger R_\mathrm{exc} \frac{\nu^\mathrm{IS}+1}{\pi r^2} \cos (2 \theta) \,,  \label{eq:t1stress}
\\
T_2^\ddagger(r,\theta) &= f^\ddagger R_\mathrm{exc} \left[ -\frac{3-\nu^\mathrm{IS}}{2} \delta (r)+\frac{\nu^\mathrm{IS}+1}{\pi r^2} \cos (4 \theta )\right]\,, \label{eq:t2stress}
\\ 
T_3^\ddagger(r,\theta) &=  -f^\ddagger R_\mathrm{exc} \frac{\nu^\mathrm{IS}+1}{\pi r^2} \sin (4 \theta)\label{eq:t3stress}
\end{align}
where $\nu^\mathrm{IS}$ is the 2D Poisson's ratio and $\delta(r)$ is the Dirac delta function; see SM \cite{Note1}, Sec.~2.1-2.2. 

To compute the barrier $\Delta F^\ddagger$ from Eqs.~\eqref{eq:t1stress} to \eqref{eq:t3stress}, one must first propose an appropriate reaction coordinate $q$. One candidate for $q$ is the second principal invariant $J_2$ of the deviatoric stress tensor $S_{ij} \equiv T_{ij}-\frac{1}{d}\delta_{ij} T_{kk}$ where $J_2  = \frac{1}{2} S_{ij} S_{ij}$. The invariant $J_2$ has precedence in solid mechanics as a criterion for plastic yield, e.g. von Mises yield criterion \cite{gurtin2010mechanics} and fully specifies the deviatoric part of the elastic strain energy corresponding to second term in Eq.~\eqref{eq:2delasticenergy}, i.e., 
\begin{equation}
\Delta F_\mathrm{d}  = \int \diff^2 \* x \frac{1}{2 G^\mathrm{IS}} \left[ (T_2)^2+(T_3)^2 \right] =  \frac{1}{2 G^\mathrm{IS}} \int \diff^d \* x \ J_2  \,; \label{eq:devpartenergy}
\end{equation}
see SM \cite{Note1}, Sec.~2.2.

Setting the reaction coordinate as $ q \equiv \sqrt{\int \diff^2 \* x \ J_2 }$, the deviatoric strain energy in Eq.~\eqref{eq:devpartenergy} corresponds to a 1D energy well given by 
\begin{equation}
\Delta F_\mathrm{d}(q) = \frac{1}{2} k q^2 \label{eq:1denergywell}
\end{equation}
where $k = 1/G^\mathrm{IS}$. Using Eq.~\eqref{eq:1denergywell}, one can compute the barrier $\Delta F^\ddagger$ by evaluating $q$ at its transition-state value $q^\ddagger$ using Eqs.~\eqref{eq:t1stress} to \eqref{eq:t3stress}, and integrating in polar coordinates with $R_\mathrm{exc}$ as a short-distance cutoff (SM \cite{Note1}, Sec.~2.2). This results in the following expression for the TST energy barrier
\begin{equation}
\Delta F^\ddagger = \frac{(f^\ddagger)^2 (1+\nu^\mathrm{IS})^2}{2 G^\mathrm{IS} \pi} \,. \label{eq:elasticbarrier}
\end{equation}
Since the elastic moduli are computable from Eqs.~\eqref{eq:elasticitytensor} to \eqref{eq:mismatchforce}, the only unknown left in Eq.~\eqref{eq:elasticbarrier} is the force magnitude $f^\ddagger$.

\section{The T1 Transition State}  \label{sec:t1transition}

To compute $f^\ddagger$, we use the concept of an eigenstrain \cite{Balluffi2012}, which is a transformation strain $\epsilon_\mathrm{c}$ inside the STZ. Eigenstrains are typically used to study the effect of inclusions in elastic solids \cite{eshelby1957determination,eshelby1959elastic}. By analyzing the elastic stresses at the boundary of the STZ core (SM \cite{Note1}, Sec.~3.1), one can write $f^\ddagger$ as a function of excitation size $R_\mathrm{exc}$, eigenstrain $\epsilon_c$, and the elastic constants as 
\begin{equation}
f^\ddagger = \frac{2 \pi R_\mathrm{exc} G^\mathrm{IS} \epsilon_\mathrm{c}}{\sqrt{2}(1+\nu^\mathrm{IS})}    \,. \label{eq:forcemag}
\end{equation} 
Using Eq.~\eqref{eq:forcemag}, the TST energy barrier in Eq.~\eqref{eq:elasticbarrier} becomes
\begin{equation}
\Delta F^\ddagger = G^\mathrm{IS} \pi  R_\mathrm{exc}^2 \epsilon_\mathrm{c}^2 \,, \label{eq:jsigmainitial}
\end{equation}
where the eigenstrain $\epsilon_\mathrm{c}$ still needs to be determined. 

Computing $\epsilon_c$ requires an understanding of how particles move and reorganize to create microscopic pure shear. Inspired by the rearrangement processes in 2D cellular networks \cite{weaire2001physics,cantat2013foams}, we propose a T1 transition event as a mechanism for inducing shear deformations corresponding to the force dipole configurations. Recall that a T1 transition involves the rearrangement of four neighboring cells in a Voronoi network. If each Voronoi cell is occupied by a particle, then a T1 transition may proceed as in Fig.~\ref{fig:t1_transition}a, which is consistent with the force-dipole configuration shown in Fig.~\ref{fig:dipoleconfig}a.

Suppose the particle configuration is triangulated so that each edge represents an elastic bond between neighbors as shown in Fig.~\ref{fig:t1_transition}b. In this representation, a T1 transition is equivalent to applying pure shear to a polygonal cell. Using the parametrization of the geometry shown in Fig.~\ref{fig:t1_transition}b, let $u^\ddagger$ be the magnitude of the displacement $\* u_1$ of node \Circled{1} leading to the transition state. Using the fact that the area of the cell does not change during a pure shear deformation, one can compute the displacements of all the nodes and the overall strain in the STZ. This then relates the eigenstrain $\epsilon_c$ to the displacement $u^\ddagger$ of node \Circled{1} given by 
\begin{equation}
\epsilon_\mathrm{c} = 2\sqrt{2}\left[\dfrac{u^\ddagger/\sigma(1+u^\ddagger/\sigma)}{1+2u^\ddagger/\sigma}\right] \,; \label{eq:e_c}
\end{equation}
see SM \cite{Note1}, Sec.~3.1 for a detailed derivation.

Note that the T1 transition produces a bond-breaking event between particle \Circled{1} and \Circled{3}. If $\tilde{u}^\ddagger$ sets the onset of this event, then its value should be constrained so that the length segments joining nodes \Circled{1} and \Circled{3} $\ell_{13}$ and nodes \Circled{2} and \Circled{4} $\ell_{24}$ must be subjected to the constraint  $\ell_{13} \leq \ell_{24}$. This constraint implies that the excitation size $R_\mathrm{exc}$ can be set to $R_\mathrm{exc}\equiv \frac{\ell_{24}}{2}$, which encompasses the transition state configuration corresponding to the T1 transition event. Using the geometry in Fig.~\ref{fig:t1_transition}b, the formula for $R_\mathrm{exc}$ can be written as (SM \cite{Note1}, Sec.~3.1)
\begin{equation}
R_\mathrm{exc} \equiv \frac{\sqrt{3}\sigma}{2(1+2u^\ddagger/\sigma)} \,. \label{eq:rexc}
\end{equation} 
Because of the constraint ($\ell_{13} \leq \ell_{24}$) imposed on $u^\ddagger$, an upper theoretical limit $u^\ddagger_\mathrm{max}$ also exists that can be solved by the condition $\ell_{13}=\ell_{24}$ yielding
\begin{equation}
u^\ddagger_\mathrm{max} = \frac{1}{2}\sigma(-1 + 3^{1/4}) \approx 0.158\sigma \,.    
\end{equation}
This theoretical limit points to very small strains that may be needed to trigger a reorganization event of the particles. It will also be useful when discussing the computational results in Sec.~\ref{sec:results}. 

Given the size of excitations and the eigenstrains as a function of $u^\ddagger$ in Eqs.~\eqref{eq:e_c} and \eqref{eq:rexc}, a final formula for the TST energy barrier in Eq.~\eqref{eq:jsigmainitial} can be obtained as 
\begin{equation}
\Delta F^\ddagger=  6 \pi G^\mathrm{IS} \sigma^2   \frac{(\tilde{u}^\ddagger)^2(1 + \tilde{u}^\ddagger)^2}{(1 + 2 \tilde{u}^\ddagger)^4} \, \label{eq:tstbarrierfinal}
\end{equation}
where $\tilde{u}^\ddagger = {u^\ddagger}/{\sigma}$ is the last remaining unknown.

\section{Relating Eigenstrain to Local Structure} \label{sec:eigenstrain}

In Sec.~\ref{sec:t1transition}, we derived a formula for the eigenstrain $\epsilon_\mathrm{c}$ as a function of a displacement variable $u^\ddagger$, which sets the onset of an elastic bond-breaking event. These bond-breaking events correspond to the reorganization of the first solvation shell, and can be characterized by the inherent state radial distribution function (RDF). 

Since glass formers are typically multi-component systems, the relevant RDF should be obtained from averaging the partial RDFs.
To that end, suppose the system has continuous poly-dispersity as considered in this work and its pair potential is written in a form $\phi(r/\sigma_{\alpha \beta})$ where $\sigma_{\alpha \beta}$ is a function of $\alpha$-th and $\beta$-th particle diameter. In this setting, an averaged RDF can be defined as 
\begin{equation}
\tilde{g}^\mathrm{IS}(\tilde{r}) = \frac{1}{N(N-1)/2} \sum_{\alpha, \beta} \tilde{g}_{\alpha \beta}^\mathrm{IS}(\tilde{r}) \label{eq:rdf}
\end{equation}
where $\tilde{r} = r/\sigma_{\alpha \beta}$ and $\tilde{g}_{\alpha \beta}^\mathrm{IS}(\tilde{r})$ is the inherent state partial RDF between $\alpha$-th and $\beta$-th particles computed from histograms of the dimensionless inherent state pairwise distance $R^{\alpha \beta}/\sigma_{\alpha \beta}$. Using Eq.~\eqref{eq:rdf}, one can compute static inherent-state properties as if they come from an effective mono-disperse system, e.g, the virial pressure in 2D can be calculated as $P^\mathrm{IS} = -\frac{\pi \rho^2}{2} \int_0^\infty \diff \tilde{r} \ \tilde{r}^2 \phi_{\tilde{r}}(\tilde{r}) \tilde{g}^\mathrm{IS}(\tilde{r})$ (see SM \cite{Note1}, Sec.~3.2 for agreement with the Irving-Kirkwood virial pressure). This implies that Eq.~\eqref{eq:rdf} provides a compact yet self-consistent picture of inherent-state local structure. 

\begin{figure}[t]
\centering
\includegraphics[height=0.6\linewidth]{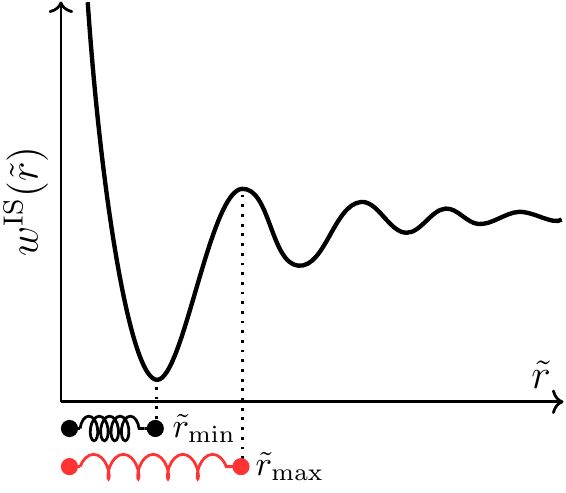}
\caption{An illustration of the inherent-state PMF showing the elastic bond at equilibrium in black, and the onset of bond-breaking event in red.}
\label{fig:pmf}
\end{figure}

A better description of the bond-breaking event can be obtained by considering the inherent-state potential of mean force (PMF) corresponding to $\tilde{g}^{\mathrm{IS}}(\tilde{r})$ defined as  $w^\mathrm{IS}(\tilde{r}) = -k_\mathrm{B} T \log \tilde{g}^\mathrm{IS}(\tilde{r})$. As illustrated in Fig.~\ref{fig:pmf}, an elastic bond can be broken by displacing a particle sitting in the first well of $w^\mathrm{IS}(\tilde{r})$ to the nearest saddle point. Denoting $\tilde{r}_\mathrm{min}$ and $\tilde{r}_\mathrm{max}$ as the locations of the first energy well and saddle point respectively, the displacement $u^\ddagger$ can be computed as 
\begin{equation}
\tilde{u}^\ddagger = \frac{1}{2}\left(\tilde{r}_\mathrm{max}-\tilde{r}_\mathrm{min}\right) \,. \label{eq:uddagger}
\end{equation}
Since $\tilde{r}_\mathrm{min}$ sets the contact distance of $\tilde{g}^\mathrm{IS}(\tilde{r})$, $\sigma$ in the T1 transition event is given by $\sigma \equiv \langle \sigma \rangle \tilde{r}_\mathrm{min}$ where $\langle \sigma \rangle $ is the average particle diameter. 

\begin{figure*}[t]
    \centering
    \includegraphics[width=\linewidth]{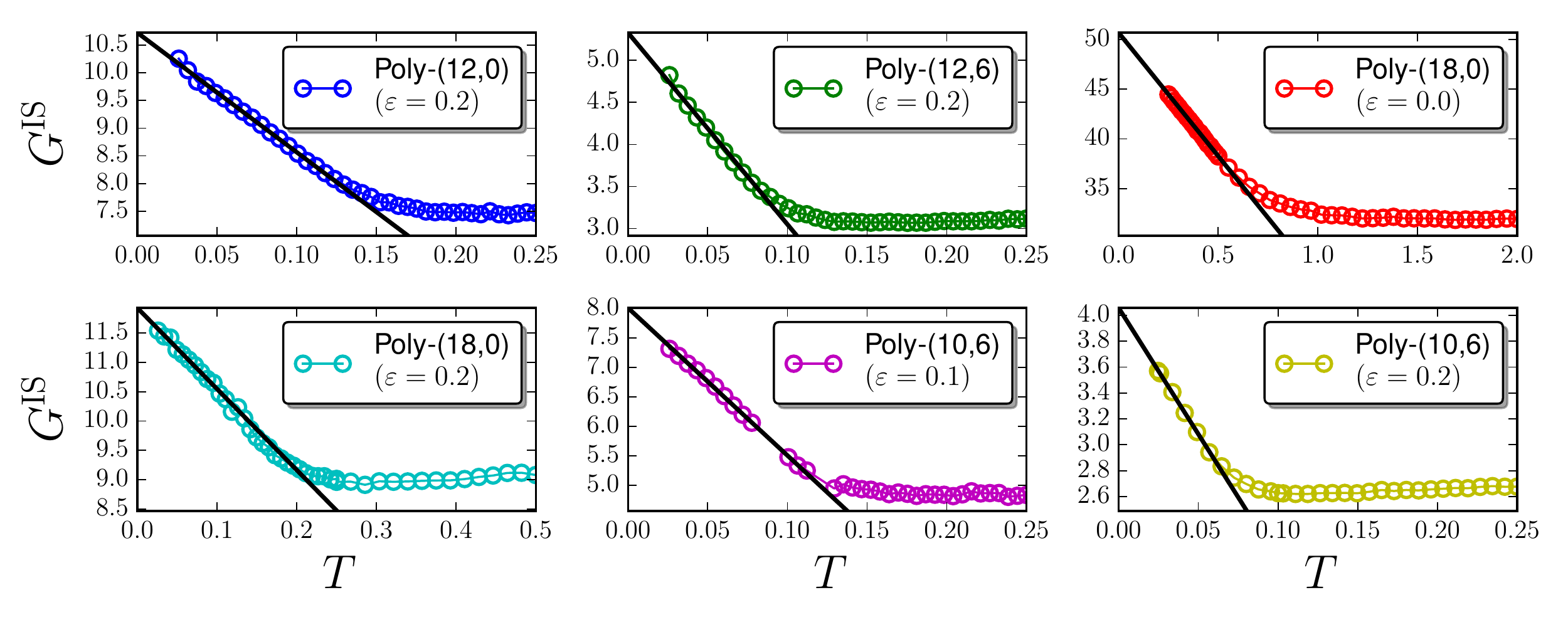}
    \caption{Plot of inherent-state shear modulus $G^\mathrm{IS}$ as a function of temperature $T$. Black line is a a linear fit to the low-$T$ portion of the data. Note that $G^\mathrm{IS}$ also exhibits a high-$T$ plateau-like regime.}
    \label{fig:allshearmodulus}
\end{figure*}

With a formula for $\tilde{u}^\ddagger$ and $\sigma$ at hand, $J_\sigma$ can be obtained as a zero-temperature limit of Eq.~\eqref{eq:tstbarrierfinal}, 
\begin{equation}
J_\sigma = \lim_{T \to 0} \left[6 \pi G^\mathrm{IS} \sigma^2   \frac{(\tilde{u}^\ddagger)^2(1 + \tilde{u}^\ddagger)^2}{(1 + 2 \tilde{u}^\ddagger)^4} \right] \label{eq:jsigmafinal}
\end{equation}
where $G^\mathrm{IS}$, $\tilde{u}^\ddagger$, and $\sigma$ are functions of temperature. However, recall from Sec.~\ref{sec:ratetheory} that the validity of Eq.~\eqref{eq:jsigmafinal} comes with the assumption of linearity in $\Delta F^\ddagger$ with respect to temperature. On the other hand, it is well-known that local structure changes very little with respect to temperature. Furthermore, we show for all the poly-disperse models considered in this work, $u^\ddagger$ is also practically independent of temperature (see SM \footnote{See Supplementary Material} Fig.~S15b). Therefore, if our assumption is correct, then the linearity of $\Delta F^\ddagger$ should arise mostly from $G^\mathrm{IS}$, which will be verified in the next section. 

\section{Results \& Discussions} \label{sec:results}

To validate the current theory, we compare its prediction for $J_\sigma$ in Eq.~\eqref{eq:jsigmafinal} with the ones computed using DF theory \cite{Keys2011} on a class of continuous poly-disperse atomistic models \cite{Ninarello2017}. The continuous poly-dispersity in these systems coupled with the Monte Carlo (MC) swap algorithm \cite{Ninarello2017} has been shown to obtain equilibrium configurations at ultra low temperatures, which are essential in calculating the elastic moduli and the ensuing barriers.

\begin{table}[t]
\caption{List of poly-disperse models and their key parameters. Density $\rho=1.01$, and system size is set to $N=32^2$ particles.} \label{tab:modellist}
\begin{ruledtabular}
\begin{tabular}{lcccc}
Model\footnote{A model with repulsive-interaction exponent $m$ and attractive-interaction exponent $n$ is named Poly-$(m,n)$}   & $m$ & $n$ & $\varepsilon$ & $\tilde{r}_\mathrm{c}$  \\ \hline
Poly-(12,0) & 12 & 0 & 0.2  & 1.25 \\
Poly-(12,6) & 12 & 6 &0.2  & 2.5  \\
\multirow{2}{*}{Poly-(18,0)} & 18 & 0 & 0.0 & 1.25  \\
     & 18 & 0 & 0.2 & 1.25 \\
\multirow{2}{*}{Poly-(10,6)} & 10 & 6 & 0.1 & 2.5         \\
     & 10 & 6 & 0.2 & 2.5    
\end{tabular}
\end{ruledtabular}
\end{table}

The poly-disperse systems are characterized by pair potentials of the form 
\begin{equation}
\phi(r/\sigma_{\alpha \beta}) = v_0 \left[\left(\frac{\sigma_{\alpha \beta}}{r}\right)^{m}-\left(\frac{\sigma_{\alpha \beta}}{r}\right)^{n} \right] + F(r/\sigma_{\alpha \beta}),
\end{equation}
for $r/\sigma_{\alpha \beta} \leq \tilde{r}_\mathrm{c}$ and zero otherwise. Here, $F(r/\sigma_{\alpha \beta})$ is an even polynomial that keeps $\phi(r/\sigma_{\alpha \beta})$ second-order continuous at the cutoff radius $\tilde{r}_\mathrm{c}$.  The parameter $\sigma_{\alpha \beta} = \frac{\sigma_\alpha +\sigma_\beta}{2}(1-\varepsilon|\sigma_\alpha-\sigma_\beta|)$, where $\varepsilon>0$ is the non-additivity parameter. The particle diameter distribution is a power-law, i.e.,  $P(\sigma) \sim 1/\sigma^3$ for $\sigma_\mathrm{min} < \sigma < \sigma_\mathrm{max}$ and zero otherwise. In Table~\ref{tab:modellist}, we list six such poly-disperse systems based on their interaction exponents $(m,n)$, $\varepsilon$, and $\tilde{r}_\mathrm{c}$. The rest of model parameters are standardized so that the reduced units of mass $m^* = 1$, length $\sigma^* = \langle \sigma \rangle =1$, and energy $\varepsilon^* = v_0 = 1$. The computational work flow involving poly-disperse models, swap MC algorithm and its implementation in HOOMD-blue  \cite{anderson2020hoomd}, sampling of inherent states, calculations of the ensemble averaged inherent state shear modulus and inherent state RDF is provided in Appendix~\ref{app:methods}.

\begin{figure*}[t]
    \centering
    \includegraphics[width=\linewidth]{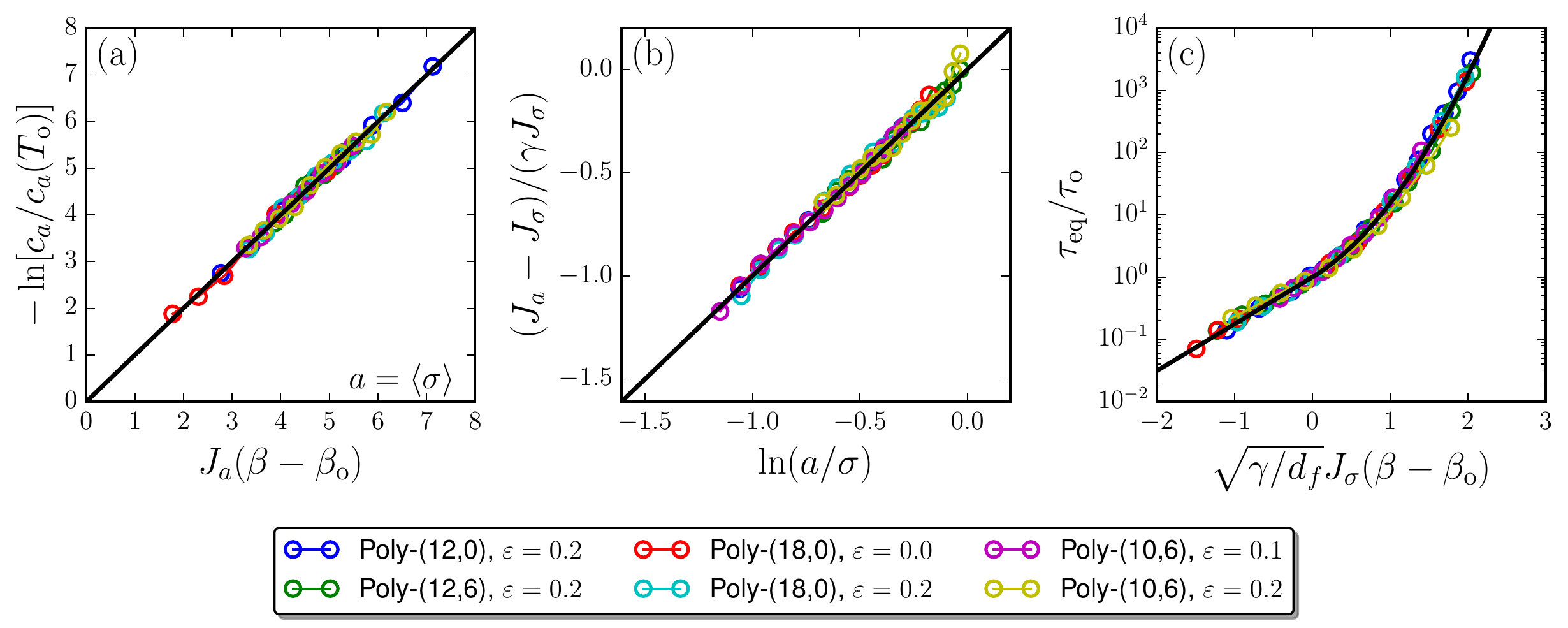}
    \caption{DF theory: (a) A plot of $-\ln[c_a]$ vs. $\beta=1/k_\mathrm{B}T$ for $a=\sigma$ showing a collapse of concentration of excitations for different poly-disperse models. (b) A plot of $J_a$ vs. $\ln(a)$ again showing a collapse for different models according to the logarithmic relation in Eq.~\eqref{eq:facilitate}. (c) An Arrhenius plot of equilibrium relaxation time $\tau_\mathrm{eq}$ where all data are collapsed according to the parabolic law (Eq.~\eqref{eq:parablaw}). Note that the $J_\sigma$ and $\gamma$ are obtained independently  from data shown in (a) and (b).}
    \label{fig:dftheory_collapsed}
\end{figure*}

\begin{table*}[t]
\caption{Table of $J_\sigma$ computed from three different methods, alongside $\gamma$ and the averaged $\tilde{u}^\ddagger$. With the exception of $\tilde{u}^\ddagger$\footnote{The reported value of $\tilde{u}^\ddagger$ is averaged over all available temperatures and uncertainty is 95\% confidence interval.}, all uncertainties are standard errors of a regression coefficient.} \label{tab:summary}
\begin{ruledtabular}
\begin{tabular}{lcccccc}
Model  & DF Theory Analysis $\gamma$ & DF Theory Analysis $J_\sigma$  & Parabolic Law Fit $J_\sigma$  & Current Theory $J_\sigma$  & $ \tilde{u}^\ddagger$ \\ \hline
Poly-(12,0), ($\varepsilon=0.2$)    & 0.177(2)     & 1.710(2)                       & 1.78(2)            & 1.77(2)        &   0.116(1)          \\
Poly-(12,6), ($\varepsilon=0.2$)   & 0.228(4)      & 0.914(2)                       & 0.913(1)            & 0.80(2)        &  0.120(1)         \\
Poly-(18,0), ($\varepsilon=0.0$)     & 0.242(3)     & 6.69(1)                       & 6.7(2)            & 10.2(6)  & 0.156(6)           \\
Poly-(18,0), ($\varepsilon=0.2$)   & 0.169(2)    & 2.034(3)                       & 2.07(2)            & 2.18(1)      &  0.124(2)           \\
Poly-(10,6), ($\varepsilon=0.1$)      & 0.185(2)  & 1.365(2)                       & 1.367(4)            & 1.56(3)      &  0.148(1)           \\
Poly-(10,6), ($\varepsilon=0.2$)   & 0.225(6)  & 0.700(2)                        & 0.669(4)          & 0.588(2)        &   0.125(2)        
\end{tabular}
\end{ruledtabular}
\end{table*}

As shown in Fig.~\ref{fig:allshearmodulus}, the shear modulus $G^\mathrm{IS}$ is linear at low temperatures for all poly-disperse models listed in Table~\ref{tab:modellist}. As mentioned previously, we find that in all of these models, the displacement $u^{\ddagger}$ computed from Eq.~\eqref{eq:uddagger} is almost constant as a function of temperature. This confirms a fundamental assumption in the current theory that $\Delta F^\ddagger(T)$ is linear at low temperatures. Given $G^\mathrm{IS}$ and the displacement $u^\ddagger$, the energy barrier $J_\sigma$ in Eq.~\eqref{eq:jsigmafinal} can be estimated by computing $\Delta F^\ddagger(T)$ in  Eq.~\eqref{eq:tstbarrierfinal}, and extracting the zero-temperature value by a linear fit. These values are summarized in Table~\ref{tab:summary}. We also note that the linear behavior of the shear modulus at low temperatures crosses over to a constant plateau behavior at high temperatures for all poly-disperse models. This observation allows us to collapse the shear modulus data of all models onto a universal curve, which then yields a value for the cross-over temperature, denoted as $T_\mathrm{p}$. The details of the data collapse procedure can be found in SM\cite{Note1} (Sec.~4.3, Fig.~S12). The physical significance behind $T_\mathrm{p}$ is left for future work.

The equilibrium relaxation time $\tau_\mathrm{eq}$ is defined such that the self-part of the intermediate scattering function $F_s\left(k=\frac{2 \pi }{\langle \sigma \rangle},t=\tau_\mathrm{eq}\right)=0.1$.
We fit the relaxation times to the parabolic form in \eqref{eq:parablaw} to obtain the onset temperature $T_\mathrm{o}$, and effective energy scale $J = \frac{\gamma}{d_f} J_\sigma$. Recall from Sec.~\ref{sec:dftheory} that $d_f \approx 1.8$ in 2D \cite{Keys2011} and thus, $J_\sigma$ can be estimated from $J$ once $\gamma$ is determined using excitation analysis from the DF theory. For more details on MD simulation protocol and parabolic-law fitting procedure, see SM \cite{Note1}, Sec.~4.5.

Following the procedure for the DF theory in Ref.~[\onlinecite{Keys2011}], we calculate the concentration of excitations $c_a(T)$ and the energy barriers $J_a$ for the poly-disperse models. The concentration of excitations $c_a(T)$ is estimated via the formula $c_a = C_a(t_a)/t_a$, where $C_a(t)$ is given by Eq.~\eqref{eq:probe_exc}, and $t_a$ is an observation time that lies within the linear regime of $C_a(t)$ (SM \cite{Note1} Fig.~S.17b).
As shown in Fig.~\ref{fig:dftheory_collapsed}a, the rate $c_a(T)$ for $a=\langle \sigma \rangle$ is Arrhenius in agreement with Eq.~\eqref{eq:excrate}. In Fig.~\ref{fig:dftheory_collapsed}b, we also see that the energy barrier $J_a$ computed from the slope of $-\ln c_a(T)$ vs. $1/T$ follows the logarithmic relation in Eq.~\eqref{eq:facilitate_exc}, and that the $J_a$ vs. $\ln a$ data can be collapsed with the fitted $J_\sigma$ and $\gamma$ values for all the models. Finally, using $J_\sigma$ and $\gamma$ obtained from the excitation analysis, we can independently estimate  $\tau_\mathrm{eq}(T)$ according to the parabolic law (Eq.~\eqref{eq:parablaw}), which is shown to be in quantitative agreement with the measured relaxation times as shown by a single universal curve in Fig.~\ref{fig:dftheory_collapsed}c. All parameters of the DF theory analysis are summarized in Table~\ref{tab:summary}. For more details on the DF theory analysis, see SM \cite{Note1}, Sec.~4.6. 

In Table~\ref{tab:summary}, we list $J_\sigma$ computed from the DF theory, parabolic-law fitting, and predictions from the current theory. Good agreement can be found between the estimates from the DF theory and the current theory, thus showing the relevance of the theory of elasticity and corresponding transition states in understanding the emergence of localized excitations. There exists one notable exception, which is Poly-(18,0) ($\varepsilon=0.0$), where the energy barrier is approximately 1.5 times the estimate from the DF theory. Interestingly, this large error coincides with having the largest displacement $\tilde{u}^\ddagger = 0.156(6)$, which is close to the theoretical limit of $\tilde{u}_\mathrm{max}^\ddagger \approx 0.158$ computed in Sec.~\ref{sec:t1transition}. Furthermore, the best agreement corresponding to Poly-(12,0) correlates with the smallest $\tilde{u}^\ddagger = 0.116(1)$. These observations indicate that the theory's accuracy may be best when the displacements needed to create an excitation are small, consistent with the usage of linear elasticity theory. It is also plausible that the nature of the reorganization events for the Poly-(18,0) ($\varepsilon=0.0$) model may not correspond to a T1 transition event requiring further investigation. 

\section{Conclusion}

In summary, we have presented a structure-based theory for understanding the origin of localized excitations as defined by the DF theory. Our theory is able to capture the energy barriers for particle displacements in the DF theory by establishing a connection with the theory of elasticity for inherent states. Note that recent work  analyzed the particle displacements and strain fields around a localized excitation event in a two-dimensional poly-disperse model \cite{chacko2021elastoplasticity}. The strain profiles emanating from our theory are consistent with the strain profiles found in Ref.~[\onlinecite{chacko2021elastoplasticity}] down to the length scale of a particle diameter, indicating our theory is appropriate towards a quantitative understanding of energy barriers for particle displacements in supercooled liquids.

Future work entails extensions of the theory to 3D, where the nature of the reorganization events and models for the transition state still remain unclear. Furthermore, it is desirable to connect the current theory to experiments. One way to achieve this is to measure the viscosity $\eta$ and the shear modulus of different low temperature glassy liquids as a function of temperature. While the viscosity measurements can be used to estimate $J = \frac{\gamma}{d_f} J_\sigma \sim J_\sigma$ with the parabolic law in Eq.~\eqref{eq:parablaw}, the zero-temperature value of the shear modulus may act as a substitute for $G^\mathrm{IS}$. If the current theory is an appropriate description of localized excitations in glass formers, then a linear correlation between the shear modulus and $J$ should be found provided that facilitated dynamics also holds.

We note that the current theory bears similarities to previous elastic models of glassy dynamics, e.g., the shoving model \cite{Dyre2006}, which utilizes theory of elasticity to describe the relaxation of glassy liquids. The crucial difference with the shoving model is that the current theory is associated with energy barriers corresponding to transitions between inherent states and not total relaxation times which includes facilitation, while the shoving model associates the energy barriers directly to total relaxation times. To this end, our theory acts primarily as a complement to the DF theory in understanding the microscopic origin of localized excitations. We also note that our theory is similar to the idea of quasi-localized modes (QLMs) \cite{kapteijns2020nonlinear,rainone2020statistical}. While our theory approximates the saddle point by the intersection of two harmonic wells with a transition state corresponding to a pair of force dipoles, the theory of QLMs appears to use the anharmonicity of the energy well. It would be interesting to establish a rigorous connection between these two approaches. 

Lastly, although the origin of facilitated dynamics remains unknown,  it is shown in previous work that a facilitation-like mechanism exists in various disordered systems. For instance, studies focusing on understanding allostery in proteins using 2D random elastic networks have shown that a localized force perturbation can trigger another force perturbation at some distance away from the original one \cite{Yan2018}. However, it remains to be seen how the current detailed theory of elasticity for localized excitations emerging from the inherent states leads to dynamical facilitation.

\section*{Supplemental Material}
The supplemental material\cite{Note1} provides the theoretical developments in greater detail, leading to the analytical formula for the energy barrier in Eq.~\eqref{eq:jsigmafinal}. It also contains details of the computational aspects of simulating the poly-disperse atomistic models and analyzing the predictions of the theory.

\begin{acknowledgments}
MRH and KKM are entirely supported by Director, Office of Science, Office of Basic Energy Sciences, of the U.S. Department of Energy under contract No. DEAC02-05CH11231. MRH also acknowledges insightful discussions with David Limmer.
\end{acknowledgments}

\section*{Data Availability}

The data that support the findings of this study are available from the corresponding author upon reasonable request.

\appendix

\section{Methods} \label{app:methods}

Simulations in Fig.~\ref{fig:allshearmodulus} were done with swap Monte Carlo (MC) \cite{Ninarello2017}  which is parallelized and implemented as a plugin \footnote{Code for parallel swap MC is available at \url{https://github.com/mandadapu-group/parallel-swap-mc}} to HOOMD-blue \cite{anderson2020hoomd}. The probability to choose swap over translational moves is $p_\mathrm{swap} = 0.2$. Inherent states corresponding to configurations equilibrated by swap MC were obtained via the FIRE algorithm \cite{Bitzek2006}. Finally, shear modulus computations were done via code developed in-house \footnote{Code for shear modulus computations is available at \url{https://github.com/mandadapu-group/pyglasstools}} and aided by the parallel eigensolver SLEPc \cite{hernandez2005slepc} to efficiently compute the pseudo-inverse of the Hessian matrix contained in the shear modulus formula (Eq.~\eqref{eq:nonaffinetensor}). Molecular dynamics (MD) simulations in Fig.~\ref{fig:dftheory_collapsed}a-c were also performed using a plugin \footnote{Code for MD simulations and energy minimization of poly-disperse models is available at \url{https://github.com/mandadapu-group/polydisperse-md}.} to HOOMD-blue. MD equilibration and production runs were done in NVT (Nose-Hoover thermostat) and NVE ensemble respectively with the timestep being $\Delta t = 0.0075$ for Fig~~\ref{fig:dftheory_collapsed}c and $\Delta t = 0.002$ for Fig~\ref{fig:dftheory_collapsed}a-b. For more details on the chosen parameters controlling MC/MD simulations, energy minimization, and shear modulus computations, see SM \cite{Note1}, Sec.~4. 

\bibliography{apssamp}

\end{document}